\documentclass[fleqn,12pt,twoside]{article}
\usepackage{espcrc1}

\usepackage{graphicx}
\usepackage[figuresright]{rotating}

\newcommand{\AmS}{{\protect\the\textfont2
  A\kern-.1667em\lower.5ex\hbox{M}\kern-.125emS}}

\title{\vspace{-2.2cm} Near-threshold $\omega$-meson production 
in proton-proton collisions\thanks{Supported by 
the Forschungszentrum-J\"{u}lich, contract No. 41445282 (COSY-058).}
:\\ With or without resonance excitations ?
}
\author{K. Tsushima$^a$, K. Nakayama$^{a,b}$\\ \vspace{1em}
$^a$Department of Physics  
and Astronomy, University of Georgia, GA 30602, USA\\
$^b$ Institut f\"{u}r Kernphysik, Forschungszentrum-J\"{u}lich, 
D-52425, J\"{u}lich, Germany
}
\begin{document}
\maketitle
\begin{abstract}
We present results for the $p p \to p p \omega$ reaction  
studied by considering two different scenarios: 
with and without the inclusion of nucleon resonance excitations. 
The recently measured angular distribution by the COSY-TOF Collaboration
at an excess energy of $Q = 173$ MeV and 
the energy dependence of the total cross section data for 
$\pi^- p \to \omega n$ are used to calibrate the model parameters.
The inclusion of nucleon resonances improves the theoretical prediction for
the energy dependence of the total cross section in $pp \to pp\omega$  
at excess energies $Q < 31$ MeV.
However, it still underestimates the data by about a factor of two, and  
remains a problem in understanding the reaction mechanism.
\end{abstract}

\section{Introduction}

\begin{figure}[htb]
\begin{minipage}[t]{75mm} 
\includegraphics[width=70mm,height=50mm]{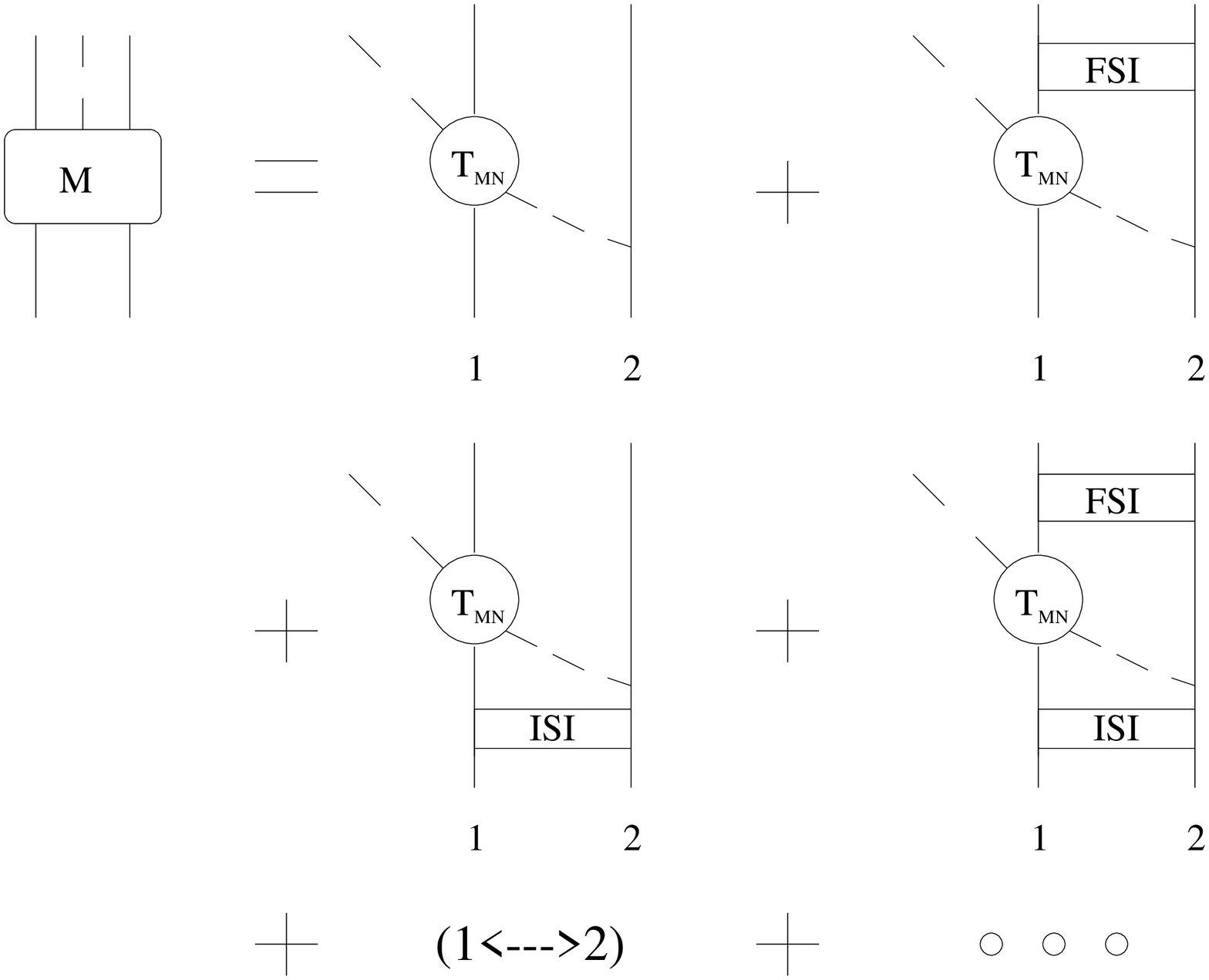}
\vspace{-0.8cm}
\caption{Amplitude $M$ for $N N \to N N \omega$. $T_{MN}$ stands 
for meson nucleon $T$ matrix 
treated with an approximation~\cite{kn}.}
\vspace{-0.4cm}
\label{fig:amplitude}
\end{minipage}
\hspace{\fill}
\begin{minipage}[t]{75mm}
\includegraphics[width=70mm,height=32mm]{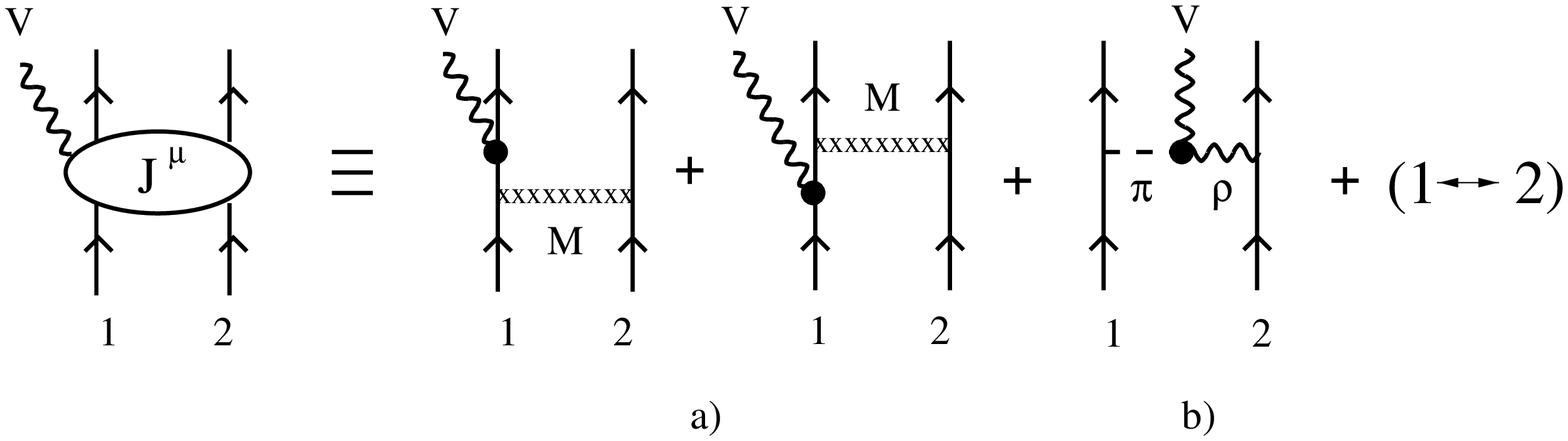}
\vspace{-0.8cm}
\caption{$\omega$ (= V) production currents, a)~nucleonic 
( and/or resonance) current,  
where, M $=\pi,\eta,\rho,\omega,\sigma,a_0$), and
b)~mesonic current.}
\vspace{-0.4cm}
\label{fig:current}
\end{minipage}
\end{figure}

One of the reasons for a renewed interest in the 
$p p \to p p \omega$ reaction  
is the recently measured $\omega$ angular distribution by the 
COSY-TOF Collaboration~\cite{COSY-TOF} at an excess energy, 
$Q = 173$ MeV. 
In addition to the existing total cross section 
data from SATURNE~\cite{Hibou} ($Q < 31$ MeV), the new data 
can be used to further constrain theoretical models.
We study the $p p \to p p \omega$ reaction  
considering two different scenarios: with and without 
the inclusion of nucleon resonance excitations~\cite{tn}.
We use a relativistic hadronic model, where the reaction amplitude
is calculated in DWBA which includes  
both the NN initial and final state interactions  
(denoted by ISI and FSI, respectively). 
The ISI is implemented 
in the on-shell approximation~\cite{tn,kn}, 
while FSI is generated using the Bonn NN potential~\cite{Bonn}.
Effects of the finite $\omega$ width are also included.
The $\omega N$ FSI is accounted for only via 
the pole diagrams (s-channel processes).
Most of the parameter values of the model are 
fixed from considerations of other 
reactions closely related to the one 
in the present study and are given in Ref.~\cite{kn}.
For further details of the model we refer to Ref.~\cite{kn}.
The $\omega$-meson production amplitude and 
its associated production currents, are depicted  
in Figs.~\ref{fig:amplitude} and~\ref{fig:current}, respectively.
The production currents consist of, a) nucleonic (and/or resonance) current
and b) mesonic current.
In section 2, we discuss the results without 
the nucleon resonance degrees of freedom,
while in section 3, we present the results with 
the resonance degrees of freedom.

\vspace{-0.3cm}

\section{Results without resonance excitations}

\begin{figure}[htb]
\begin{minipage}[t]{75mm}
\includegraphics[height=75mm,width=50mm,angle=-90]{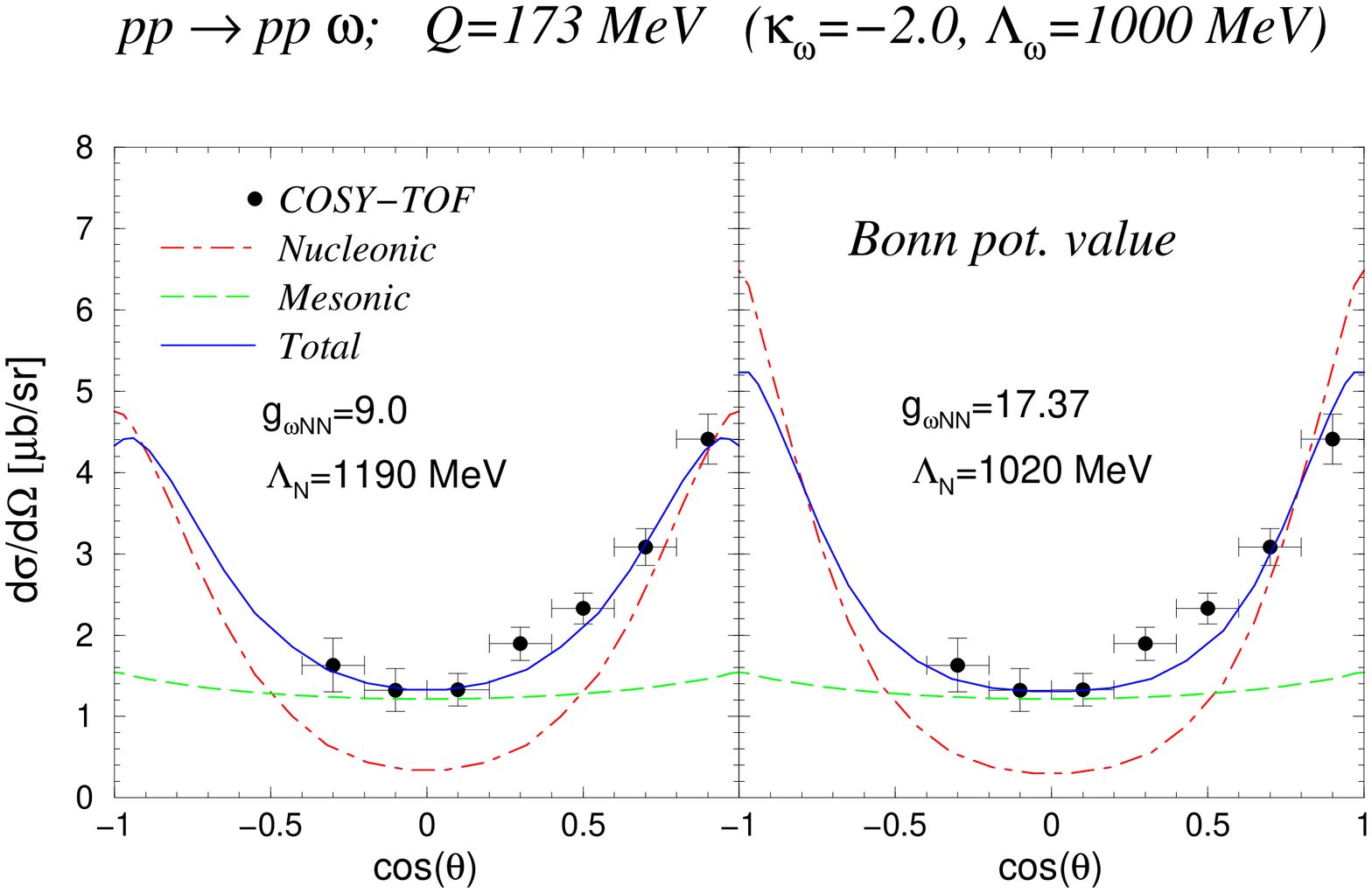}
\vspace{-0.8cm}
\caption{Calculated angular distribution with no resonances.
The left (right) panel is
the best fit (result with the Bonn potential
value for $g_{\omega NN}$). }
\vspace{-0.8cm}
\label{fig:angular1}
\end{minipage}
\hspace{\fill}
\begin{minipage}[t]{75mm}
\includegraphics[height=75mm,width=50mm,angle=-90]{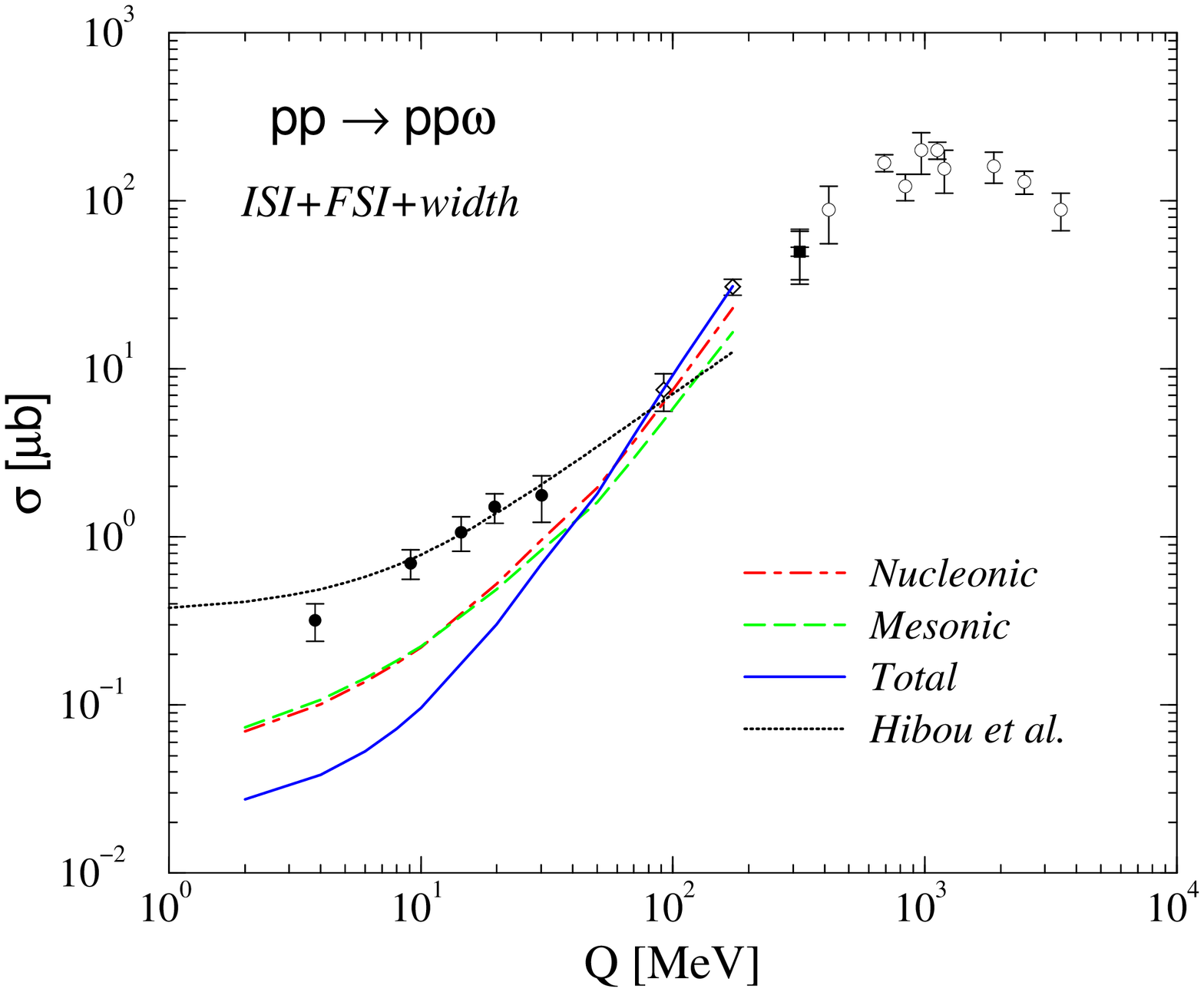}
\vspace{-0.8cm}
\caption{Energy dependence of the total cross section. "{\it Hibou et al.}"
stands for the result from Refs.~\cite{Hibou,Wilkin}.}
\vspace{-0.8cm}
\label{fig:cross1}
\end{minipage}
\end{figure}

One of the two free parameters in the nucleonic current,  
the tensor to vector coupling constant ratio, $\kappa_\omega \equiv f_\omega/g_\omega$,
at the $\omega NN$ vertex is determined by fitting the angular distribution data at 
$Q = 173$ MeV \cite{COSY-TOF}; the shape of the angular distribution is found
to be quite sensitive to the value of $\kappa_\omega$~\cite{JJHF}. Also, 
the cutoff
parameter, $\Lambda_N$, in the form factor at the $\omega NN$ vertex is 
adjusted concomitantly to reproduce the total cross section at $Q = 173$ MeV. 
Using the value, $g_{\omega NN} = + 9.0$,  
associated with the physical $\omega$ 
(not with the virtual $\omega$ exchanged), 
we obtain the value of $\kappa_\omega = -2.0$ to best reproduce
the angular distribution as shown in Fig.~\ref{fig:angular1}
(left panel).
(The value quoted in Ref.~\cite{JJHF} should read
$g_{\omega NN} = + 9.0$.)
The shape of the angular distribution is quite insensitive to the 
values of $g_{\omega NN}$, as shown 
in Fig.~\ref{fig:angular1},  
where the result in the right panel is obtained by using a value of 
$(g_{\omega NN})^2/4\pi = (17.37)^2/4\pi = 24$, 
which corresponds approximately to the value 
used in the Bonn NN potential~\cite{Bonn}.

The energy dependence of the total cross section calculated 
using the values of $\kappa_\omega = -2.0$  and 
$g_{\omega NN} = + 9.0$, is shown in Fig.~\ref{fig:cross1}  
together with the result of Ref.~\cite{Wilkin} used in the analysis 
of the SATURNE data~\cite{Hibou,Wilkin}.
The present result substantially underestimates the SATURNE 
data~\cite{Hibou} for $Q < 31$ MeV.
The reason for this discrepancy is that, as we shall show later, 
we probably overestimate too much the mesonic 
current contribution at $Q = 173$ MeV. 
This results in substantial underestimation of 
the cross section at near threshold 
due to the much more efficient destructive interference between 
the nucleonic and mesonic current contributions as the excess energy decreases.

Although currently there is no definite experimental
evidence of a $\omega$-meson
coupling to a nucleon resonance, it would be natural to expect some resonance
current contributions, especially, at nucleon incident energies involved in the
production of $\omega$ in NN collisions.
The reduction of the mesonic current at high excess energies may then be
compensated by the nucleon resonance current contribution.
We discuss such a scenario next.

\section{Results with resonance excitations}

We now include nucleon resonance degrees of freedom (resonance current) in
the model. The criteria for selecting relevant resonances are:
(i) those appreciably decay to $N + \gamma$  
to use the vector meson dominance assumption 
to produce the $\omega$,
(ii) those mass distributions are around  
$(m_N + m_\omega)$, to be able to maximally contribute near threshold,
(iii) select a minimum number of resonances 
to avoid introducing many new parameters, 
(iv) those that can describe consistently the $\pi^- p \to \omega n$ reaction.
As a result, we have selected resonances $S_{11}(1535), P_{11}(1710),
D_{13}(1700)$ and $P_{13}(1720)$.

\begin{figure}[htb]
\begin{minipage}[t]{75mm}
\includegraphics[height=75mm,width=45mm,angle=-90]{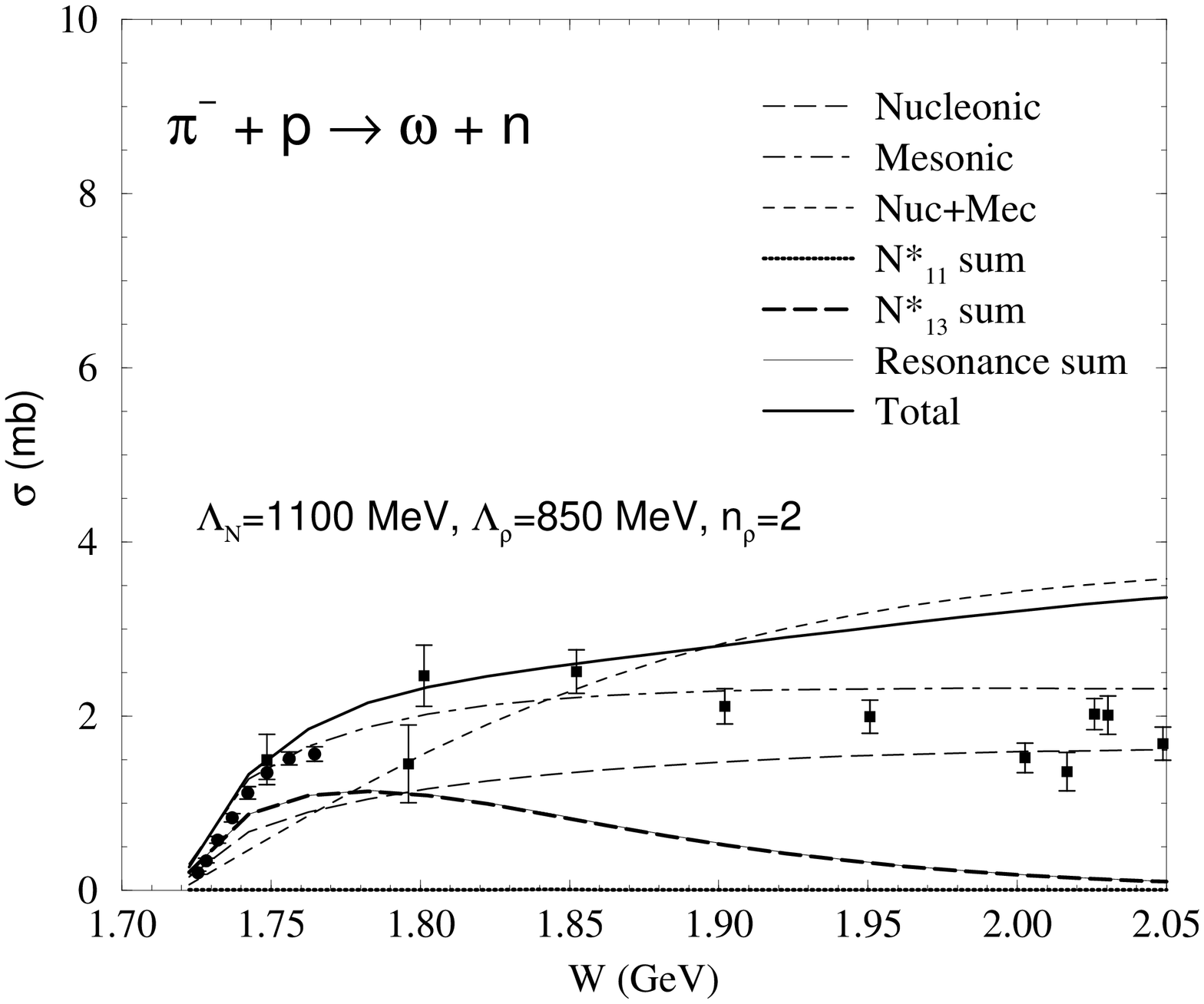}
\vspace{-1.0cm}
\caption{Energy dependence of the total cross section for 
$\pi^- p \to \omega n$.}
\vspace{-0.7cm}
\label{fig:piN}
\end{minipage}
\hspace{\fill}
\begin{minipage}[t]{75mm}
\includegraphics[height=75mm,width=45mm,angle=-90]{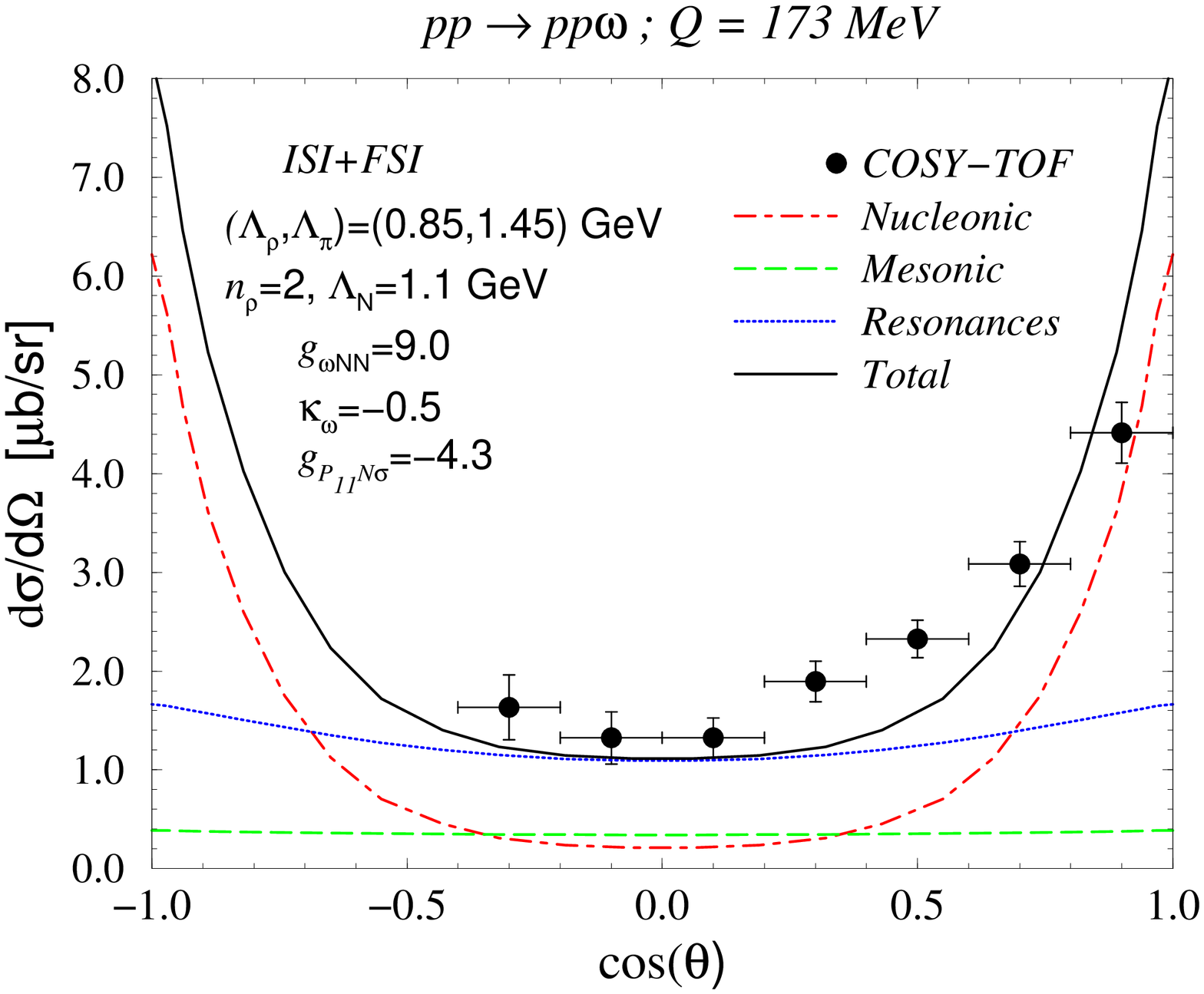}
\vspace{-1.0cm}
\caption{Calculated angular distributions including  
four nucleon resonances.}
\vspace{-0.7cm}
\label{fig:angular2}
\end{minipage}
\end{figure}

In Fig.~\ref{fig:piN} we show the calculated energy dependence 
of the total cross section for $\pi^- p \to \omega n$.
We note that at the excess energy of $Q = 173$ MeV in $p p \to p p \omega$, the
center-of-mass energy of the subsystem $\pi^- p \to \omega n$ will
reach a maximum value of $W \simeq 1.9$ GeV.
At lower $W$, $D_{13}(1700)$ and $P_{13}(1720)$ 
contributions are dominant, but not $S_{11}(1535)$ nor $P_{11}(1710)$. 
Also, without the inclusion of the resonances, it is very 
difficult to reproduce the near-threshold behavior 
of the $\pi^- p \to \omega n$ total cross section 
within a reasonable set of parameters.
With the inclusion of the resonances, we need somewhat stronger form factor for 
the $\omega \rho \pi$ vertex. In particular, the part of the form factor which
account for the off-shellness of the $\rho$ meson is of dipole form 
with the cutoff parameter 
$\Lambda_\rho = 850$ MeV~\cite{kn}.
A cutoff parameter value of $\Lambda_N = 1.1$ GeV was also determined
at the $\omega NN$ vertex. 
In addition, because $P_{11}(1710)$ has an appreciable decay branch 
to $N + 2\pi$, we simulated this by a $\sigma$-meson, where 
the value for $g_{P_{11}N\sigma}$ is adjusted to reproduce the
$pp \to pp\omega$ total cross section data at $Q = 173$ MeV.
Thus, the contribution from 
$P_{11}(1710)$ should be regarded also as taking into account the 
other possible resonance contributions not included explicitly in our model.
We obtained a value of $g_{P_{11}N\sigma} = -4.3$; 
results for another  
possible value, $g_{P_{11}N\sigma} = 4.8$, are not shown, 
because the calculated energy dependence of the total cross section  
is worse, although the angular distribution is better 
described at $Q = 173$ MeV.
Here, the value of $\kappa_\omega$ 
was set to be $-0.5$, more in line with the estimates from other 
studies. With the value of $\kappa_\omega = -2.0$ obtained in the previous
it would be very difficult to describe consistently 
the NN scattering data~\cite{Johann}.

\begin{figure}[htb]
\begin{minipage}[t]{75mm}
\includegraphics[height=75mm,width=45mm,angle=-90]{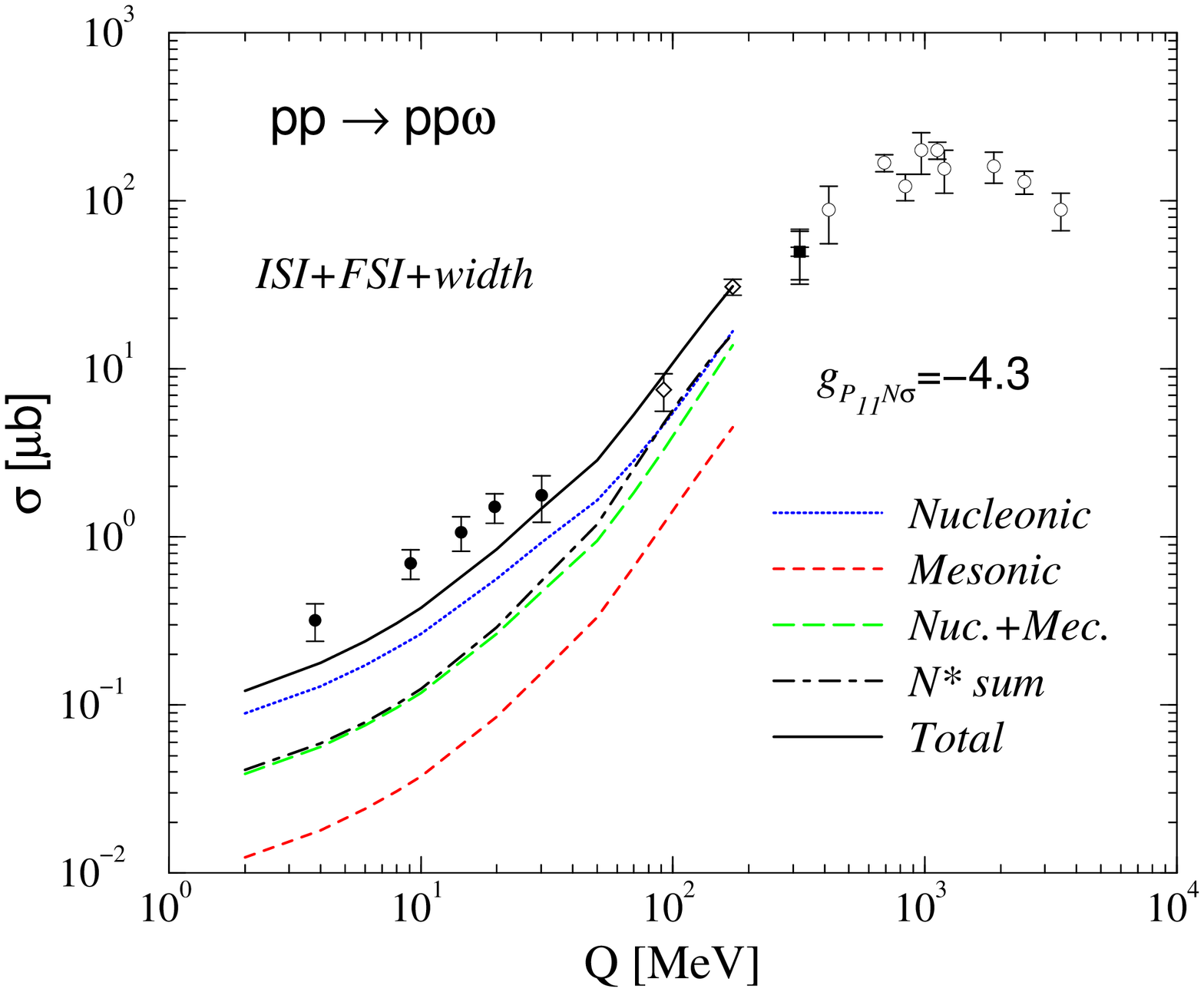}
\vspace{-1.0cm}
\caption{Energy dependence of the total cross section.}
\vspace{-0.6cm}
\label{fig:cross2}
\end{minipage}
\hspace{\fill}
\begin{minipage}[t]{75mm}
\includegraphics[height=75mm,width=45mm,angle=-90]{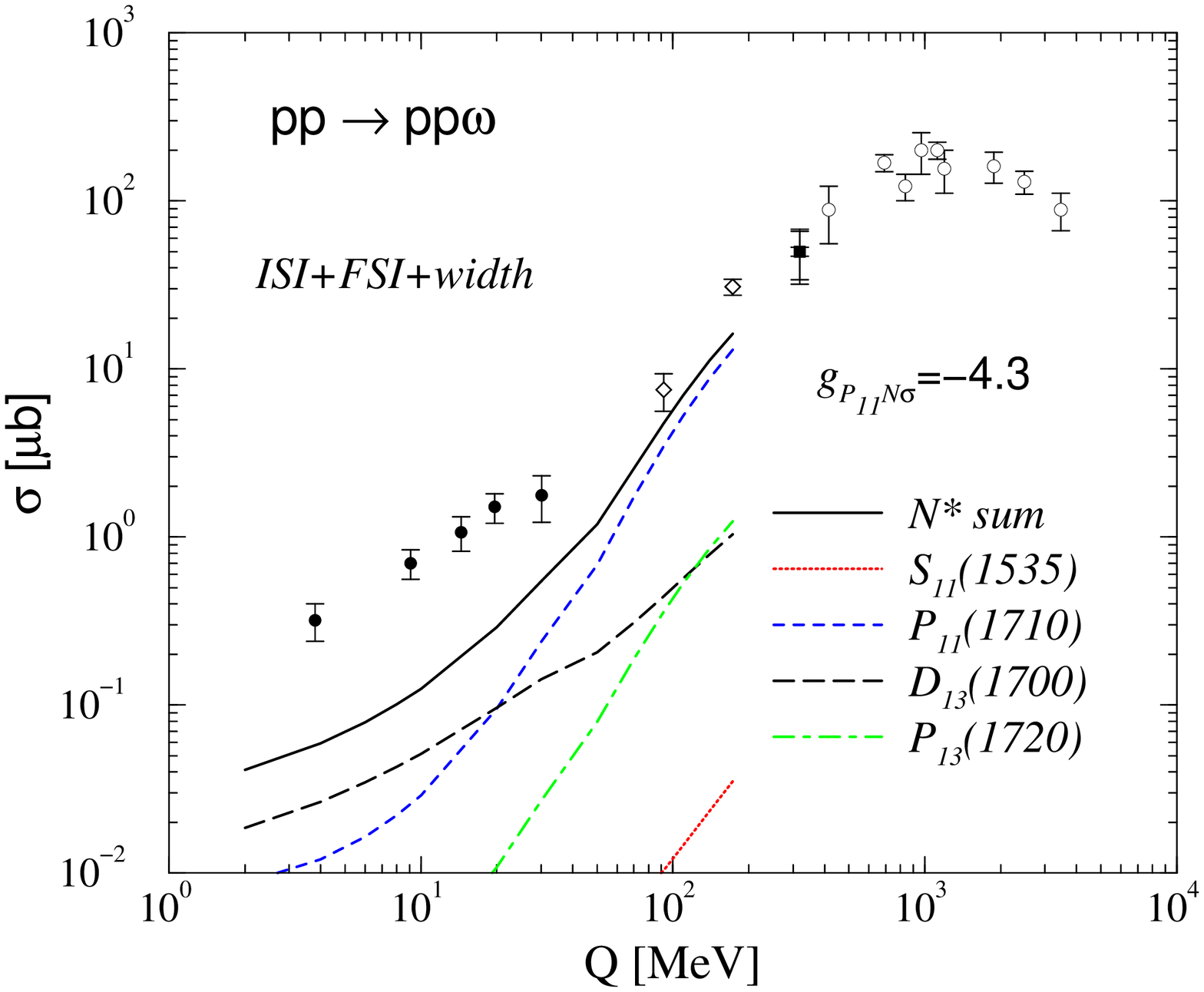}
\vspace{-1.0cm}
\caption{Decomposition of the resonance contributions.}
\vspace{-0.6cm}
\label{fig:resonance}
\end{minipage}
\end{figure}

In Fig.~\ref{fig:cross2} we show the 
calculated energy dependence of the total 
cross section for $pp \to pp\omega$.
The result is greatly improved   
compared to that without the inclusion of any 
resonances in the previous section. 
However, it still underestimates 
the SATURNE data~\cite{Hibou} ($Q < 31$ MeV) 
by about a factor of two. 
Fig.~\ref{fig:resonance} shows the decomposition of the 
resonance contribution. 
At near threshold the dominant contribution comes from $D_{13}(1700)$, 
while at higher excess energy, the 
dominant contribution comes from $P_{11}(1710)$, which was negligible 
in $\pi^- p \to \omega n$.
If we want to be consistent with both the $\pi^- p \to \omega n$ and 
$p p \to p p \omega$ reactions, it seems necessary 
to include at least $P_{11}(1710),D_{13}(1700)$ and $P_{13}(1720)$ 
resonances in the present approach.


\section{Summary}

We have reported our results for the $p p \to p p \omega$ reaction  
studied by considering two different scenarios: with and without  
the inclusion of nucleon resonance excitations.
The results show that the energy dependence of the total cross section 
$Q < 173$ MeV is apparently better described by the inclusion of 
resonance excitations, which is also consistent with the 
$\pi^- p \to \omega n$ reaction. However, it still underestimates the data 
in the $Q < 31$ MeV region by about a factor of two, and remains a problem 
in understanding the reaction mechanism. 
\vspace{-0.5em}


\end{document}